# Single-shot 3D motion picture camera with a dense point cloud

### Florian Willomitzer* and Gerd Häusler

*Institute of Optics, Information and Photonics, University Erlangen-Nuremberg, Staudtstr. 7/B2, 91058 Erlangen, Germany*
*[florian.willomitzer@fau.de](mailto:florian.willomitzer@fau.de)*

**Abstract:** We introduce a method and a 3D-camera for single-shot 3D shape measurement, with unprecedented features: The 3D-camera does not rely on pattern codification and acquires object surfaces at the theoretical limit of the information efficiency: Up to *30%* of the available camera pixels display independent (not interpolated) 3D points. The 3D-camera is based on triangulation with two properly positioned cameras and a projected multi-line pattern, in combination with algorithms that solve the ambiguity problem. The projected static line pattern enables 3D-acquisition of fast processes and the take of 3D-motion-pictures. The depth resolution is at its physical limit, defined by electronic noise and speckle noise. The requisite low cost technology is simple.

## 1. Introduction

Although highly desired, there is -surprisingly- no optical 3D-sensor that would allow for the single-shot acquisition of 3D-motion pictures with a dense point cloud. Of course, there are approaches for real time 3D-data acquisition [1-6], but they are either multi-shot or the 3D point density leaves room for improvement. We introduce a 3D-camera that combines *single shot ability* with a *dense 3D point cloud* and *high precision*. The 3D camera does not exploit color- or spatial encoding.

Obviously, the acquisition of dense 3D-point clouds from one single camera exposure is difficult. We will discuss the physical and information-theoretical limits and introduce a novel 3D-camera that is working at those limits. It displays the best possible 3D-point cloud density, which is about *15% - 30%* of the available camera pixels. The implemented sensor with a *1 Megapixel* camera delivers *160.000 - 300.000* 3D-points in each single shot. The single shot ability ( = one or several images, taken at the same time) allows for the very fast acquisition of 3D data, by flash exposure. It allows as well for the take of 3D-motion-pictures, where each camera frame includes the full field 3D-information, enabling a posteriori choice of the viewpoint.

To our knowledge, such a 3D-camera is not available yet. Among the wide spectrum of optical 3D-sensors there are sensors with high depth resolution; there are other sensors that deliver a dense point cloud; and there are (a few) sensors that allow for single shot acquisition of sparse (!) data. What are the obstacles to make a 3D-camera that comprises all three features?

The major keyword is the term *"dense point cloud"*. We could -naively- demand that each of the $N_{pix}$ camera pixels delivers a 3D-point, completely independent from its neighbors. But to avoid aliasing, we have to make sure that the image at the camera chip satisfies the sampling theorem, so a certain correlation between neighbor points is unavoidable. This is what we have to remember when we will later talk about "point cloud density" or "information efficiency": *100%* is impossible, it contradicts linear systems theory. (And indeed, all 3D-sensors display artifacts at sharp edges, where the sampling theorem is violated). In our case, satisfying the sampling theorem is as well necessary, as we exploit subpixel interpolation for high distance precision, as depicted in Fig. 1(a).

Keeping in mind that *100%* efficiency is impossible, we nevertheless will rate the information efficiency $\eta$ of a single-shot sensor by $\eta = N_{3D}/N_{pix}$ were $N_{3D}$ is number of independent 3D-pixels. For $\eta<1$, the more so for $\eta<<1$, only a "pseudo dense" surface reconstruction is possible, commonly realized by a posteriori interpolation. To make the discussion less abstract for the remaining part of our paper, we will assume that the discussed sensors display a space-bandwidth of *1 Megapixel* (*1000×1000 pixels*). An efficiency of, say *30%* will give us *300.000* independent 3D-points.

A small efficiency (or low density point cloud) implies a reduced lateral resolution of the sensor. Looking closer at the single shot solutions, for example by Artec [3] or Kinect 1 [2], it turns out that they indeed lack high lateral resolution, although the absence of fine details in the 3D-data often remains concealed by interpolation and high resolution texture mapping. The reason is that any kind of triangulation requires the identification of corresponding points, may this be for classical stereo or for the methods above. The encoding devours space-bandwidth that is lost for high lateral resolution. In [3], for example, the width of the projected stripes is encoded piecewise along the stripe direction. In [2], a pseudo-random pattern of dots is projected. Classical stereo exploits "natural" salient spacious "features". Is this fundamental? The bad news is: *"yes"*.

For better understanding let us start with the paradigm principle that delivers a "dense" point cloud, however after at least *three* exposures: the so called "fringe projection triangulation", sometimes called "phase measuring triangulation" (PMT) [7]. This principle is a local method - each pixel delivers information independent from its neighbor pixels (within the limits of the sampling theorem). As mentioned, fringe projection is not a single-shot principle. Three exposures are required, because the local distance for each pixel has to be deciphered from the *three* unknowns, ambient light, object reflectivity, and fringe phase, individually for each camera pixel. This is impossible with only one exposure, as these data are encrypted by only one intensity signal. There are single-shot workarounds such as single sideband demodulation [8], which however demands for a spatial bandwidth of the object smaller than *1/3* of the otherwise allowed bandwidth. A similar approach is the so called "spatial phase shifting". Obviously, to buy speed we have to put certain space-bandwidth on the counter. These considerations may explain why the magic number *"3"* will come across more often, in the further discussion.

So far about the multi-shot sensor with virtually *100%* efficiency. At the other end of the spectrum of sensors is the "perfect" single shot sensor principle: light sectioning triangulation [9]. Instead of fringes, only one narrow line is projected onto the object. Along this line a perfect 3D-profile can be calculated from one camera image. Note that light sectioning is not perfectly local, as the subpixel line positioning exploits at least three neighbored pixels as depicted in Fig. 1(a). Of course light sectioning with one line displays a very small efficiency $\eta \approx 0,1\%$ for our *1 Megapixel* camera example. The next obvious question is, to what extent can we improve $\eta$ for light sectioning and what does it cost?

There is already a workaround, called "Flying Triangulation" [10,11], which is a single-shot sensor with about *10* projected lines. The data from each exposure are sparse ($\eta \approx 1\%$), But the gaps between measured lines can be filled within seconds by (on-line) registration of many exposures, while the sensor is guided (even by hand) around the object. Flying Triangulation delivers dense high quality data, within seconds, but it demands for rigid objects: speaking or moving people cannot be acquired. We follow this way of thinking and ask how many lines are possible, to get a significant point cloud density within each shot.

Only for a moment, we will neglect the (profound) problem to uniquely identify each line, and estimate the maximum possible number of lines: to localize each line with subpixel accuracy, the lines must satisfy the sampling theorem and there must be sufficient space between the lines.

For subpixel interpolation, the linewidth must be wider (but not much wider, to avoid overlap) than *4p* (with pixel pitch *p*). With a half width of a little more than *2p*, precise subpixel interpolation is ensured. These numbers are in agreement with experimental experience. We have to note as well that the line image at the three evaluated pixels (Fig. 1(a)) must not be disturbed by strong variation of the height or texture. This tells us again, that independent data within a close neighborhood are not possible.

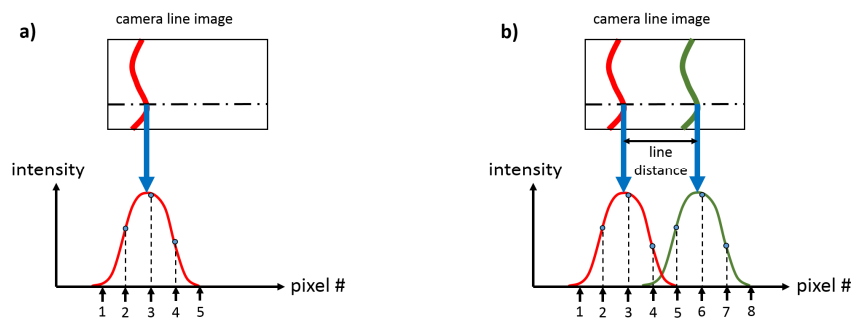

Fig. 1: (a) Nyquist sampling allows for precise sub-pixel line localization and high distance precision. (b) The minimum distance between projected lines is three times the pixel pitch.

Figure 1(b) reveals that the distance between two lines must be at least *3p*, for low crosstalk. Fine details between the lines are not resolved, and, with proper band limitation, should not occur.

A camera with $N_x$ pixels in *x*-direction will allow for a maximum of $L \approx N_x/3$ lines, and the sensor can acquire $N_{pix}/3$ valid 3D-points, within one single shot, giving an efficiency of $\eta \approx 33\%$ or *330.000* 3D-pixel with a *1 Megapixel* camera. Why are we not surprised to find again the magic number *3*?

For a big triangulation angle and highly tilted areas, the camera may see a smaller line distance, due to perspective contraction. This will require a larger line distance to be projected. With, say, a line distance of *6p*, the (pessimistic) efficiency will be $\eta \approx 16\%$, with *160.000* 3D-points of our *1 Megapixel* camera. We will demonstrate that this efficiency is realistic and can be technically achieved without extreme requirements for calibration and mechanical stability.

Now to the crucial question: how to correctly identify (to "indicate"), e.g., *330* lines, or more modest, *160* lines? It was discussed that this formidable ambiguity problem cannot be solved by spatial encoding of the lines - simply because there is not enough space at the (*1 Megapixel*) chip.

There is, however, a different and principally perfect solution, by exploiting color as another modality. The so called "rainbow sensors" [12] and [13] use a projected color-spectrum to encode the distance via triangulation. A color camera decodes the shape from the hue of each pixel. Principally, color encoded triangulation may have an efficiency $\eta = 100\%$. We notice that there are three color channels of a three-chip color camera which buy us faster measurement by more space-bandwidth (=more pixels). Although the concept of color triangulation is long known, it is not yet well established - possibly because it prevents color texture acquisition and it is difficult to project bright saturated spectra. However, new broad band fiber lasers may serve as light sources for rainbow sensors, in the future.

From the *"three chip"* camera it is a small step to ask if we can replace the three (red, green, blue) modalities by using a couple of synchronized black-and-white cameras. With several cameras, the identification of each projected "line" or "pixel" may become much easier, if not unique.

The idea to use many cameras was already presented by [14], about *10* years ago, and a principal solution was demonstrated. The authors project a pattern with binary stripes onto the object, while *C* cameras are required to distinguish $2^C$ depths. With a proper choice of the triangulation angles (in exponential sequence), each stripe can uniquely be identified.

This was (to our knowledge) the first "proof of principle" for a single-shot 3D-camera with a potentially dense point cloud. It reveals that unique triangulation can be achieved by several images synchronously taken at the same moment, instead of a temporal sequence of images. But note again: this method may improve the density of the 3D-point cloud, but it does not reach the *100%* efficiency of phase measuring triangulation[1].

How does the concept of [14] match our considerations above? For a setup comparable to ours (*160* lines, *1000* distinguishable distances) the method described in [14] would require a multitude of cameras. We will demonstrate that *two* cameras are sufficient to measure up to *300.000* 3D-pixels. Moreover, due to proper subpixel interpolation, the precision of our method is limited only by coherent noise or electronic noise.

---

[1] To compare single shot and multi-shot principles, which exploit $N_{frame}$ temporal exposures for the evaluation of one 3D dataset, it is reasonable to divide our information efficiency $\eta$ by $N_{frame}$. This "frame efficiency" would be again *33%* for phase measuring triangulation. Common PMT-sensors are less efficient, as they exploit many more exposures for better precision and uniqueness.

## 2. Single-shot 3D movie camera with uni-directional lines

As discussed in the previous section, a single-shot principle does not supply sufficient information to provide data with *100%* efficiency (disregarding rainbow triangulation). If, in addition, pattern encoding comes into play, the efficiency will be even less.

Our single-shot camera is based on *multi-line light sectioning*, without any encoding to indicate the lines. The decoding is done just by combining the images of two properly positioned cameras.

We will describe two different approaches, with different projected patterns. The first approach exploits a projected pattern of straight, narrow (binary) lines (*160* lines for a *1 Megapixel* camera). The object is observed by *two* cameras, from different triangulation angles.

How to manage the necessary "indexing" of that many lines? The corresponding ambiguity problem is discussed in a previous paper [15]. As the novel solution is an extension of the earlier results, these are briefly summarized:

For common multi-line triangulation, the achievable line density $L/\Delta x$ ($L$ is the number of projected lines, $\Delta x$ is the width of field) is related with the triangulation angle $\theta$ and the unique measurement depth $\Delta z$ by:

$$\frac{L}{\Delta x} \leq \frac{1}{\Delta z \cdot \tan \theta} \qquad (1)$$

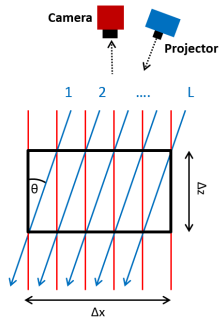

Fig. 2: The achievable number of lines $L$ depends on the triangulation angle θ and the unique measurement depth $\Delta z$.

A violation of Eq. (1) results in false 3D-data, observed as outliers (see Fig. 4(e)). Reference [15] explains how these outliers can be detected and corrected by exploiting the data of a second camera, positioned at a second angle of triangulation. The basic idea is to (virtually) project the data of the first camera back onto the camera chip of the second camera. The correctly evaluated data can be detected as they necessarily coincide at the camera chip (but, commonly, not the outliers). However, with increasing line density, more outliers from one camera coincide by chance with data from the second camera and the achievable (unique) line density is only moderate. For sufficient density, registration of several frames is required.

Now, we will introduce an effective improvement of the "back-projection" idea, to overcome this limitation. We will demonstrate that a thoughtfully designed optics for the illumination and the observation, in combination with moderately sophisticated software can solve the problem[2].

---

[2] Our experience is, by the way, that properly designed optics is often preferable to a posteriori image processing: optics serves as a powerful source encoder [16] that may enable otherwise impossible tasks or at least facilitates the decoding problems.

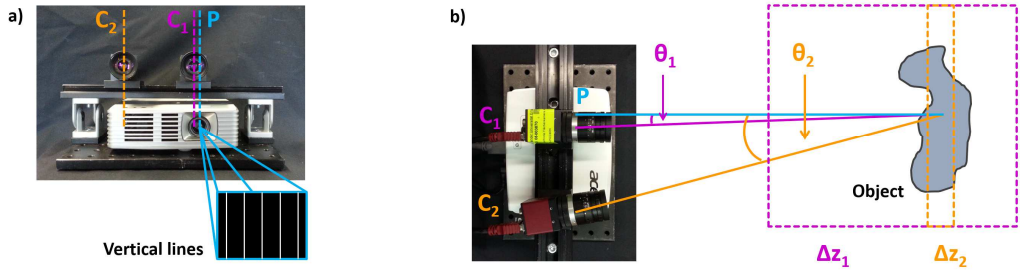

Fig. 3: Setup, comprising a projector *P* that projects a static binary line pattern, and two cameras $C_1$ and $C_2$. For a vertical line pattern, the horizontal distance between the nodal points of projector and camera define the relating triangulation angles $\theta_1$ and $\theta_2$. (a) front view of the setup. (b) View from top, sketching relating angles and measurement volumes.

The basic idea of the novel modification is as follows: again, *two* cameras and *one* projector are used (see Fig. 3). The first camera $C_1$ and the projector *P* create a triangulation sensor with a very small triangulation angle $\theta_1$. This first sensor delivers noisy but unique data within the required measurement volume $\Delta z_1$, according to Eq. (1). The data are noisy, as the precision is $\sim 1/\sin\theta_1$. The second camera $C_2$ and the projector create a second triangulation sensor with a bigger triangulation angle $\theta_2$. This second sensor delivers more precise data, but with ambiguity. As both sensors look at the same projected lines, the first sensor can "tell" the second sensor the correct index of each line, via the same back-projection mechanism as described in [15]. With the proper choice of the triangulation angles, there is no accidental overlap of false and correct data.

According to Fig. 4, a narrow line pattern with *L* lines is projected onto the object (Fig. 4(a)). The object is observed by two cameras ($C_1$ and $C_2$) under two triangulation angles $\theta_1$ and $\theta_2$. The small $\theta_1$ enables the required large measurement depth $\Delta z_1$ according to Eq. (1) and Fig. 3(b). The triangulation angle $\theta_2$ is chosen considerably larger for a high precision of the measured distance.

The observed line images deviate from a straight line, depending on the triangulation angle. The lines seen by $C_1$ are nearly straight and can be easily indexed, as shown in

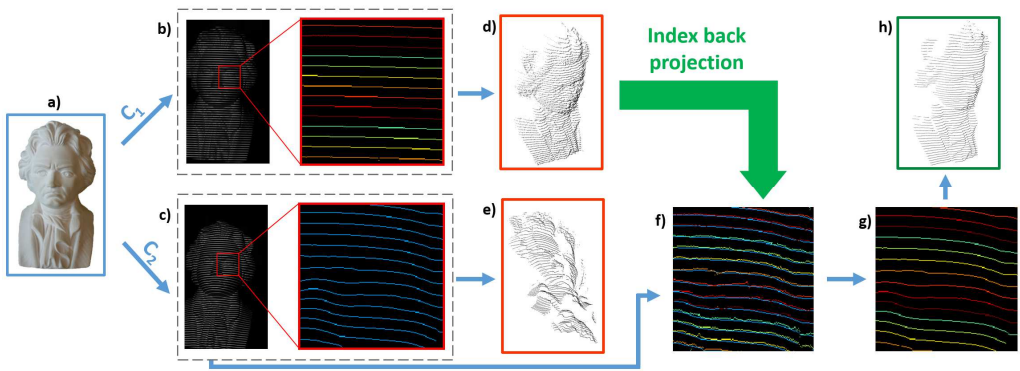

Fig. 4: Unique indexing by combining two camera images: (a): object; (b) and (c): images of the object with two cameras seen from different triangulation angles. (known indices in (b) are color coded). (d) and (e): 3D data, directly calculated from (b) and (c). (f) noisy, correctly indexed 3D data from (d) (color coded), back-projected onto the chip of $C_2$, together with the line image of $C_2$. (g) correctly indexed lines of $C_2$, assigned from the indices delivered by $C_1$ (color coded). (h) final 3D model evaluated from (g) with correct indices and low noise.

Fig. 4(b). In the sketch, the index is illustrated by a color code. The directly calculated 3D model (Fig. 4(d)) displays correct indexing but high noise. Direct 3D point calculation from $C_2$ produces low noise but errors by ambiguity (Fig. 4(e)). To solve this problem, both sets of information are merged: the points of Fig. 4(d) including their index information are back-projected onto the chip of $C_2$. Implying a precise calibration [17] and considering geometric constraints, the back-projections overlap with the line signal (Fig. 4(f)). Eventually, the back-projected line indices of $C_1$ are assigned to the corresponding lines on the chip of $C_2$ (Fig. 4(g)), which enables a correct evaluation with high precision (Fig. 4(h)).

As the reader might guess from Fig. 4(f), $\theta_1$ and $\theta_2$ cannot be chosen independently. Noise has to be taken into account. The correct index can uniquely be assigned as far as the back-projected noisy line images from $C_1$ do not crosstalk with the neighbor line images on $C_2$. More precisely: the back-projected (noisy) lines should not overlap with other than the corresponding lines seen by the second sensor. This is the case if Eq. (2) is satisfied

$$\frac{L}{\Delta x'} \leq \frac{1}{\delta x'} \cdot \frac{\sin \theta_1}{\sin \theta_2}, \qquad (2)$$

where $\delta x'$ represents the chip sided uncertainty of the line localization, due to noise, and $\Delta x'$ is the chip width. For human face data, the major source of noise is electronic camera noise. Speckle noise is largely suppressed by volume scattering within the skin. Our data display $\delta x' \approx 0,1$ *pixel*. With $\theta_2 \approx 9°$, and a field view of *350mm*, we estimate a distance uncertainty of $\delta z$ better than *200μm*. Figure 5(c) illustrates the low noise in the measured data. We add that the stand-off distance is *550mm* and the achieved unique measurement depth $\Delta z$ is about *100mm*, with *160* projected lines.

Several measurements were performed with the proposed setup. To demonstrate the robustness of the principle against varying object texture and reflectivity, "natural" objects (like human faces) are measured (instead of the omnipresent white plaster busts with Lambertian scattering).

Figure 5 displays raw data from a single-shot measurement of a human face with a line density of *160 lines/field*, acquired with a *~1 Megapixel* camera (*1024x768 pixels*). This relates to an information efficiency or "3D-point density" of $\eta \approx 16\%$. Figure 5(b) shows the acquired 3D data of the object (Fig. 5(a)). Note that all different perspectives are extracted from the same single video frame. Black-and-white texture information is already included in the 3D data. In Fig. 5(c) a close look at the data illustrates the good precision. We emphasize that the displayed 3D models are *not* post-processed. No interpolation or smoothing is applied. Each displayed 3D point is measured independently from its neighbors.

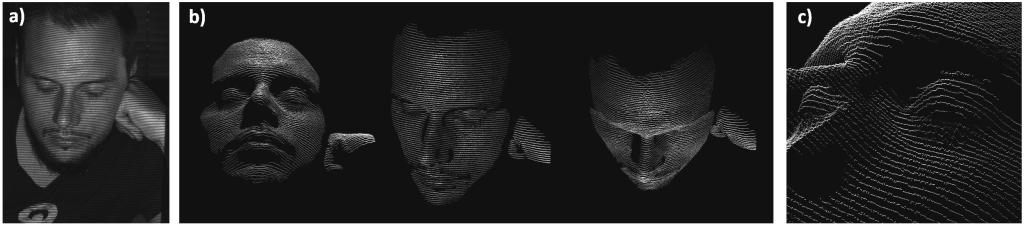

Fig. 5: Raw data (no post-processing) from a single-shot measurement with a line density of *160 lines/field*. (a) Camera image of human face with projected lines. (b) 3D model from different perspectives, evaluated from one single video frame. (c) close look at the 3D data, illustrating the low distance noise.

With the described single-shot method, object surfaces can be measured with considerable density and a precision only limited by (coherent or electronic) noise [18]. The time required for a single measurement is as short as the exposure time for one single camera frame. Just a static binary pattern (slide, chrome on glass) is projected which does not require an electronically controlled pattern generator. So the 3D data can be acquired in milliseconds or microseconds, depending on the available amount of light. As each camera frame includes a 3D model, 3D-motion pictures can be acquired. Examples are shown in the next section and in [19].

**3. With crossed lines towards higher point density**

In the introduction we estimate the maximum number of lines in practice cautiously to ~*160 lines* for a *1 Megapixel* camera. It becomes obvious from Fig. 6(a) that it might be very difficult to implement more lines (the reader might zoom in).

But there is another option: our first approach can indeed be upgraded by the projection of *crossed lines*. Figure 6(b) displays the *original* pattern to be projected, with *160* vertical lines and *100* horizontal lines, according to aspect ratio of the projector.

After image acquisition, the two line directions are first identified, isolated and eventually, separately evaluated. Principally, a second pair of cameras could be added to evaluate the second perpendicular line direction. But there is a more simple and cost effective solution, requiring again *only two cameras* instead of four: As depicted in Fig. 3(a), only the distance of the camera and projector perpendicular to a line direction defines the triangulation angle: for a crossed line pattern, we can generate two different triangulation angles for each camera. The resulting setup (see Fig. 7) has four triangulation angles - one large and one small angle for each line direction. With only two cameras we create four triangulation sensors.

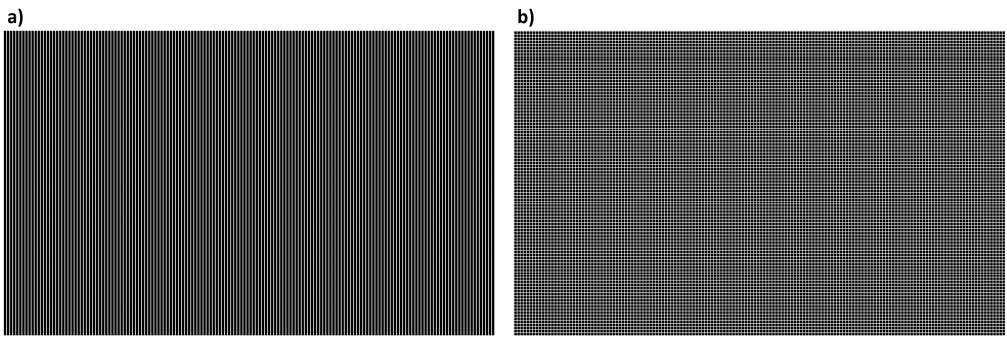

Fig. 6: Original patterns to be projected onto the object, to demonstrate the density of 3D points.
(a) uni-directional line pattern (see sec. 2). (b) crossed lines for higher point density (see sec. 3).
The reader might zoom in to resolve the patterns.

Principally, this can be done even with more cameras and line directions. Such a setup with $C$ cameras and $D$ line directions could produce $C \times D$ triangulation sub-systems. The setup shown in Fig. 7 ($C=D=2$) is acquiring nearly *300.000* 3D-points by a single frame of a *1 Megapixel* camera (crossing points are counted only once). Again, it turns out that a proper optical setup makes things easy.

Where is the compromise compared to the camera with only one line direction? What is the cost of the increased number of 3D-pixels? The identification of the line direction requires some (not too serious) restriction of the surface shape: to distinguish different directions, a small line segment has to be visible, which requires some neighborhood and a certain "smoothness" of the surface. This means, not all measured 3D-points are completely independent from the neighborhood anymore. We add that it is advantageous to increase the intensity of the projected pattern at the crossing points. So the line position can still be evaluated in both directions. However, this decreases the signal-to-noise ratio at the other line segments, which decreases the precision. The related total efficiency of $\eta \approx 30\%$ is already close to the maximum possible frame-efficiency of fringe projection systems.

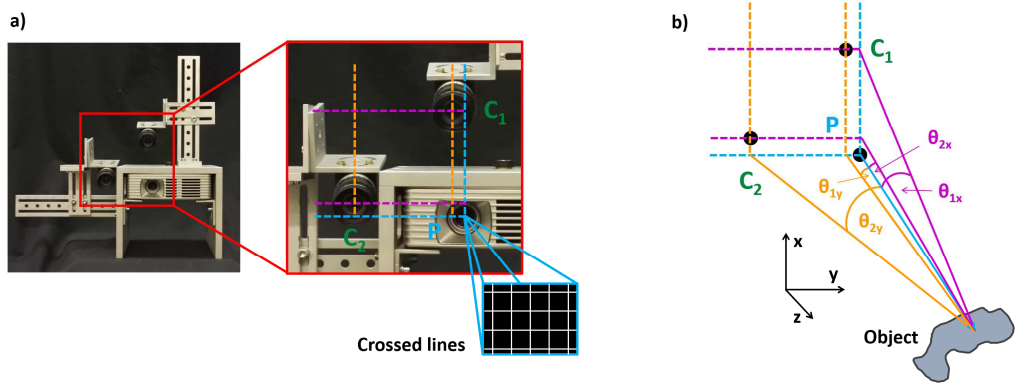

Fig. 7: Setup with crossed line projection: Two cameras $C_1$ and $C_2$ and the projector $P$ produce *four* independent triangulation angles $\theta_{1x}$, $\theta_{1y}$, $\theta_{2x}$ and $\theta_{2y}$. Each line direction is evaluated separately. (a) front view of the setup. (b) perspective sketch with triangulation angles.

Figure 8 displays frames (again with unprocessed not interpolated raw data) of a 3D-movie, acquired with the setup shown in Fig. 7. The different perspectives are again calculated from the same video frame. Figure 8(b) illustrates the low noise.

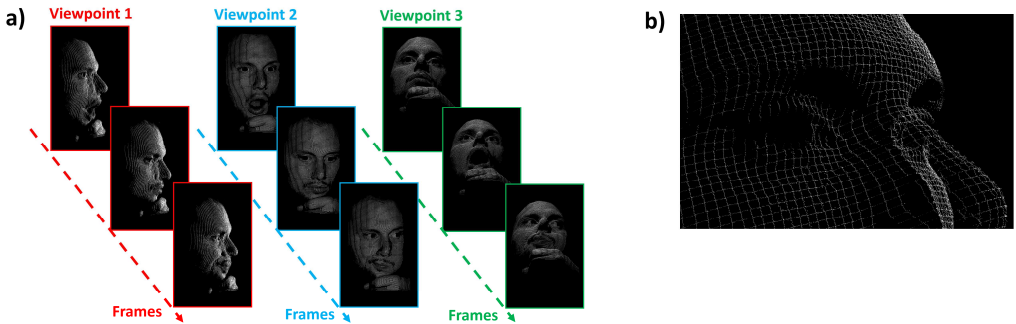

Fig. 8: (a) single frames of a 3D movie of a talking human face acquired with cross line single shot triangulation. The different perspectives (viewpoints 1, 2, 3) are all taken from the same corresponding video frame. (b) close look at the 3D data, illustrating relatively low noise.

### 4. Summary

A single-shot 3D-camera concept and device is presented, for the acquisition of up to *300.000* 3D-points, within each single camera frame of a *1 Megapixel* video camera. This number is close to the limit, as theoretical estimations reveal. The camera exploits triangulation with a pattern of *160* uni-directional lines or with a pattern of crossed lines with the same pitch. The fundamental problem of unique line identification is solved by combining the images of two cameras, one with a very small triangulation angle and the other a big triangulation angle.

The 3D-camera is technically simple. Special care is given to the proper geometry of optics and illumination and to obey the sampling theorem. The precision is limited only by coherent or electronic noise. It is in the range of *1/1000* of the distance measuring range. The performance of the 3D-camera is demonstrated by 3D-videos of human faces. More videos can be seen on our YouTube channel [19].

**5. A retrospective aha-experience**

"Why can't we buy a single-shot 3D-sensor that delivers a dense point cloud?" [20] The answer bothered our research group for some time: there is simply not enough information from one video pixel. As there is, depressingly, no perfect single-shot sensor, we at least tried to find out the limits of the achievable point density. Comparing phase measuring triangulation, a single sideband workaround and simple counting of (band limited) lines, the consistent results appear convincing: *33%* density should be possible.

Unfortunately, the estimation of the limit does not comprise an obvious recipe for a sensor that indeed reaches this limit. Really? Not before finishing the project (as so often) we had to notice that the solution was always plain on the table. We want to share these retrospective insights with the reader.

Those who ever saw an image plane hologram (or an interferogram) will notice the striking resemblance with the images of Fig. 9. And indeed, the phase of the lines ("fringes") encodes the depth, as a hologram or an interferogram do. The image of Fig. 9 (a) could even be optically reconstructed as a hologram, by a laser. After eliminating the base band and the second diffraction image by single side band filtering (easily performed by Fourier optics), a perfect reconstruction of the object shape is possible. (Of course, the data have to be re-scaled to compensate for the small wavelength of light).

Why doesn't this work for the image of Fig. 9 (b)? Due to the big triangulation angle, the phase modulation is much larger than $2\pi$ and a unique object reconstruction is impossible (there are attempts by so called "unwrapping"). In terms of the already discussed bandwidth constraints, a holographic reconstruction (=single sideband demodulation) is impossible, as the object band cross talks with the base band.

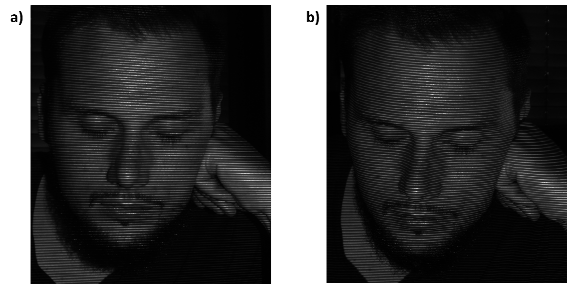

Fig. 9: Line images seen by $C_1$ (small $\theta_1$) and $C_2$ (large $\theta_2$). (a) The line image of $C_1$ displays a unique "phase distortion" ($<2\pi$) can be considered as a perfect image plane hologram of the object surface. (b) The line image of $C_2$ displays a large phase modulation ($>2\pi$), which will not allow for simple unique decoding.

After all, the first sensor, with the extremely small triangulation angle is the key component: it serves as a "phase modulation compressor" or a "bandwidth compressor" enabling the acquisition of objects with a large depth variation. To say it in the words of an opticist: the first sensor in combination with the second sensor exploits information concepts of holography or multi-wavelength interferometry *for rough, macroscopic objects.*

Eventually, back to depression: we have to accept the bandwidth constraints of single-shot principles. But state of the art video cameras supply a plethora of pixels. Looking at Fig. 6, or Figs. 5 and 8, even the *1 Megapixel* camera is good for more than hundred thousand pixels, or for 3D-metrology with significant lateral resolution. Cameras with many more pixels are available and, principally, full-HD quality should not be impossible.

Acknowledgement: This paper is essentially about principles and limits, as the sensor works with simple technology. However, the sensor would not work at the limits, without precise calibration. We want to acknowledge the invaluable contributions of Florian Schiffers, who was involved in many fruitful discussions and, specifically in the calibration [17].